\documentclass[11pt]{article}
\begin{document}

\noindent
\Large
{\bf CLASSICAL MECHANICS WITHOUT DETERMINISM}
\normalsize
\vspace*{1cm}

\noindent
{\bf Hrvoje Nikoli\'c}

\vspace*{0.5cm}
\noindent
{\it
Theoretical Physics Division \\
Rudjer Bo\v{s}kovi\'{c} Institute \\
P.O.B. 180, HR-10002 Zagreb, Croatia \\
E-mail: hrvoje@thphys.irb.hr}

\vspace*{2cm}

\noindent
Classical statistical particle mechanics 
in the configuration space 
can be represented by a nonlinear 
Schr\"odinger equation. Even without assuming the existence 
of deterministic particle trajectories, the resulting quantum-like 
statistical interpretation is sufficient to predict all 
measurable results of classical mechanics. 
In the classical case, the wave function that satisfies 
a linear equation is positive, which is the
main source of the fundamental difference 
between classical and quantum mechanics. 
\vspace*{0.5cm}


\section{INTRODUCTION}

It is widely accepted that quantum mechanics (QM) is a fundamentally 
indeterministic theory, the predictions of which are 
inherently probabilistic. Unfortunately, this makes QM conceptually 
very different from the fundamentally deterministic classical mechanics (CM).
It is possible to interpret QM in a deterministic way similar 
to that in CM, by adopting the Bohmian interpretation 
\cite{bohm,bohmprl,bohmrep,hol}.
In this interpretation, particles are assumed to have definite
properties (positions, velocities, and continuous deterministic
trajectories) even when
they are not measured, which is why these properties are usually regarded
as hidden variables.
However, since this interpretation
seems to have the same directly observable predictions as the conventional
interpretation, this interpretation is
usually ignored by the physics community, mainly because
this interpretation is technically more complicated than
the usual interpretation, without clearly leading to new measurable
predictions. (The claim that there are no new measurable predictions
seems to be true in nonrelativistic QM, but not necessarily in
relativistic QM \cite{nik}.) 
With such a positivistic scientific reasoning, 
if a determininistic theory (such as 
the Bohmian interpretation of QM) can be replaced with a simpler 
indeterministic theory (such as QM with the usual interpretation) 
that leads to the same measurable predictions, then it is the 
latter indeterministic theory that should be adopted as  
a more fundamental theory.  
However, even in classical physics,
all measurements are always plagued by finite errors of measured
quantities. Consequently, even in classical physics, 
all {\em measurable} predictions are necessarily statistical.   
This suggests the following interesting question:
Is it possible to replace
the usual deterministic approach 
to CM with a {\em simpler} indeterministic approach 
similar to QM, such that all measurable predictions of the 
conventional deterministic CM are preserved?
By representing classical statistical mechanics by a nonlinear 
Schr\"odinger equation \cite{schil,ros,hol,ghose} and by adopting a certain
QM-motivated definition of ``simplicity",  
in this paper we argue that it is possible! 
More specifically, we argue that this nonlinear Schr\"odinger equation
with the appropriate purely probabilistic interpretation is sufficient
to predict all measurable results of CM.
The usual deterministic particle trajectories of CM 
play a role of unmeasured hidden variables for CM, analogous 
to the Bohmian hidden variables for QM. 
Thus, the same reasoning leading 
to the conclusion that QM is a fundamentally indeterministic theory 
may be used to argue that CM is also a fundamentally
indeterministic theory.

The paper is organized as follows.
In Sec.~\ref{SCH}, we present
a brief overview of QM and the corresponding
Bohmian interpretation, derive the nonlinear Schr\"odinger equation 
for CM, and discuss those interpretational aspects of the latter 
that are not directly related to the theory of measurement.
In Sec.~\ref{MEAS}, we present a general discussion of the problem 
of measurement in nonlinear QM and  
discuss the implications on the nonlinear classical Schr\"odinger equation.
In Sec.~\ref{CLAS}, we discuss how the standard classical laws of 
physics (classical statistics and classical trajectories) emerge from
from the quantum-like interpretation of the nonlinear classical Schr\"odinger
equation. 
A discussion of our results is presented in Sec.~\ref{DISC}.

\section{SCHR\"ODINGER EQUATIONS FOR QM AND CM}
\label{SCH}

\subsection{QM and the Bohmian interpretation} 

A nonrelativistic particle is described by the 
Schr\"odinger equation 
\begin{equation}\label{sch}
\left[ \frac{\hat{{\bf p}}^2}{2m} + V({\bf x},t) \right] \psi({\bf x},t)
=i\hbar\partial_t\psi({\bf x},t),
\end{equation} 
where 
\begin{equation}\label{p}
\hat{{\bf p}}=-i\hbar\nabla .
\end{equation}
We write $\psi$ in the polar form
\begin{equation}\label{psi}
\psi({\bf x},t)=R({\bf x},t)e^{iS({\bf x},t)/\hbar} ,
\end{equation}
where $R$ and $S$ are real functions and
\begin{equation}\label{R>0}
R({\bf x},t)\geq 0.
\end{equation}
The complex equation (\ref{sch}) is equivalent to a set of two real
equations
\begin{equation}\label{HJQ}
\frac{(\nabla S)^2}{2m}+V+Q=-\partial_tS,
\end{equation}
\begin{equation}\label{cons}
\partial_t\rho + \nabla\left( \rho\frac{\nabla S}{m} \right) =0,
\end{equation}
where
\begin{equation}\label{Q}
Q\equiv -\frac{\hbar^2}{2m}\frac{\nabla^2 R}{R},
\end{equation}
and $\rho\equiv R^2$. Eq.~(\ref{cons}) is the conservation equation 
that provides the consistency of the interpretation of $\rho$ 
as the probability density. Eq.~(\ref{HJQ}) is similar to the 
classical Hamilton-Jacobi equation, differing from it only in 
containing the additional $Q$-term. 

The similarity of (\ref{HJQ})
with the Hamilton-Jacobi equation suggests
the Bohmian interpretation 
consisting in the assumption that the particle has a definite trajectory 
${\bf x}(t)$ satisfying 
\begin{equation}\label{bohm}
\frac{d {\bf x}}{dt}=\frac{\nabla S}{m}.
\end{equation}
Eq.~(\ref{bohm}) is identical to the analogous equation in the classical 
Hamilton-Jacobi theory. Eqs.~(\ref{bohm}) and (\ref{cons}) imply 
that particles in a statistical ensemble with the initial 
probability distribution $\rho({\bf x}, t_0)$ will be distributed
according to $\rho({\bf x}, t)$ at {\em any} time $t$. 
According to the Bohmian 
interpretation, the dynamics is fundamentally deterministic, 
while all QM uncertainties 
emerge from the lack of knowledge of the actual initial particle position 
${\bf x}(t_0)$. The statistical predictions of the Bohmian interpretation
are equivalent to those of the conventional interpretation. 
Since the additional 
assumption (\ref{bohm}) does not lead to new measurable predictions, 
most physicists ignore the Bohmian interpretation and adopt the 
simpler, conventional interpretation that does not contain the additional 
equation (\ref{bohm}).    

\subsection{Classical Schr\"odinger equation}

Now consider a classical statistical ensemble. The probability distribution
$\rho$ of particle positions satisfies the continuity 
equation (\ref{cons}). 
By introducing $R\equiv\sqrt{\rho}$, this equation 
can be written in terms of a 
linear operator acting on $R$ as
\begin{equation}\label{lin}
\left[ \partial_t +\left( \frac{\nabla S}{m} \right) \nabla
+ \left( \frac{\nabla^2 S}{2m} \right) \right] R =0.
\end{equation}
Instead of (\ref{HJQ}), we have the classical Hamilton-Jacobi equation
\begin{equation}\label{HJ}
\frac{(\nabla S)^2}{2m}+V=-\partial_tS.
\end{equation}
By defining a new quantity $\psi$ as in (\ref{psi}), 
one finds that the classical equations (\ref{lin}) and (\ref{HJ}) are 
equivalent to a classical analog of the Schr\"odinger equation 
\cite{schil,ros}
\begin{equation}\label{schcl}
\left[ \frac{\hat{{\bf p}}^2}{2m} + V- Q \right] \psi =i\hbar\partial_t\psi,
\end{equation}
where $Q$ is defined as in (\ref{Q}) and $\hat{{\bf p}}$ as in (\ref{p}).
In contrast to the ordinary Schr\"odinger equation, Eq.~(\ref{schcl}) 
contains an additional $Q$-term that makes the equation nonlinear in 
$\psi$. 

The set of real equations (\ref{cons}) and (\ref{HJ}) is equivalent
to the complex equation (\ref{schcl}). However, while the 
Planck constant $\hbar$ does not appear in (\ref{cons}) and (\ref{HJ}), 
it {\em does} appear in the classical equation (\ref{schcl}). 
Does it mean that the physical content of (\ref{schcl}) does not 
depend on the value of $\hbar$? This is almost true, but not quite!
The requirement that $R$ and $S$ should be single-valued functions
does {\em not} allways guarantee that $\psi$ will also be a 
single-valued function. The additional requirement that 
$\psi$ should be single-valued may lead to 
physical results that depend on the value of $\hbar$. 
As an example, consider a particle moving circularly with the 
angular momentum $L_{\varphi}=mvr$, where $r$ is the radius 
of the circle and $v$ is the velocity. 
The solution of (\ref{HJ}) is 
$S(\varphi,t)=L_{\varphi}\varphi-Et$, where 
$\varphi$ is the angle and $E$ is the energy.
The requirement that $\psi$ should be single-valued 
means that $e^{iS(\varphi,t)/\hbar}$
should be single-valued, which reduces to the requirement
that $e^{iL_{\varphi}\varphi/\hbar}$ should be single-valued.
However, this requirement is nothing but the Bohr quantization condition
\cite{schil} 
\begin{equation}
mvr=n\hbar 
\end{equation}
of the so-called old quantum theory, for which it is known 
that correctly predicts the spectrum of the hydrogen atom.
Thus, the requirement that the classical wave function $\psi$ should
be single-valued improves standard CM, in the sense that it also 
incorporates the old quantum theory. In fact, the Bohr-Sommerfeld 
postulates of the old quantum theory could have been derived 
from the classical Schr\"odinger equation as above.     

However,
in CM, one does not consider Eqs.~(\ref{lin}) and 
(\ref{HJ}) as giving the complete description of classical systems.
Instead, one assumes that particles have definite trajectories given 
by Eq.~(\ref{bohm}). On the other hand, the success of QM (without 
the Bohmian interpretation) suggests the following interesting question:
Is it possible to recover all measurable predictions of CM just
by viewing (\ref{schcl}) as a quantum 
nonlinear Schr\"odinger equation,
without assuming the additional ``hidden-variable" 
equation (\ref{bohm})? In the subsequent sections we argue 
that it is possible. 

\section{MEASUREMENT}
\label{MEAS}

\subsection{Measurement in nonlinear QM}

The problem of measurement in nonlinear QM 
is much more delicate than that in ordinary linear QM.
For example, if one assumes that the collapse of the wave function 
is something that really happens during measurements, then, 
in contrast to linear QM, the EPR correlations {\em can} be used
to transmit information {\em instantaneously} \cite{gis,pol,cza}.
This suggests that the concept of wave-function collapse might not
be a valid concept in nonlinear generalizations of QM. Instead, 
the appearance of an {\em effective} collapse in linear QM can be 
qualitatively explained as follows. Assume that a unit-norm solution of 
a linear or a nonlinear Schr\"odinger equation can be written as a sum 
\begin{equation}\label{wf}
\psi({\bf x},t)=\sum_a c_a\psi_a({\bf x},t), 
\end{equation}
where $\psi_a$ are orthonormal functions. In the linear case, we  
also assume that the functions $c_a\psi_a$ themselves, as well as the
functions $\psi_a$, are also solutions. On the other hand, 
it is typical of the nonlinear case that 
the functions $c_a\psi_a$ and $\psi_a$ are not solutions.
Now assume that, by performing a measurement, we obtain knowledge  
that the actual state of the system is described by the 
$c_a\psi_a$ component of (\ref{wf}). To calculate the subsequent
properties of the system at later times, it is therefore sufficient to know 
only the $c_a\psi_a$ component. However, since $c_a\psi_a$ is not 
a solution for a nonlinear case, in order to calculate 
the time evolution of $c_a\psi_a$, one actually has to calculate the   
{\em whole} solution (\ref{wf}), despite the fact that only one 
of the components is the ``active" one. It is only in the linear case 
that $c_a\psi_a$ and $\psi_a$ are solutions themselves, so that
$c_a\psi_a$ evolves
independently of the other components and 
that $\psi_a$ evolves independently of the value 
of the constant $c_a$. In the linear case, this is why 
the wave function appears 
as if it ``collapsed" to $\psi_a$. 
 
An interpretation of QM that does not introduce a true collapse
of the wave function is the many-worlds interpretation \cite{witt}.
However, in the nonlinear case, such an interpretation leads
to a communication between branches of the wave function that belong
to different worlds \cite{pol}. But if there is a communication 
between the branches, then it is not really meaningful 
to say that they belong 
to different worlds. Thus, it is more reasonable to adopt 
an interpretation in which the words 
``many-worlds" are not taken so literally (see, e.g., Ref.~\cite{teg}).

The measurement in QM can also be described in a more concrete way 
as follows. Assume that $\psi_a({\bf x},t)$ in (\ref{wf}) are 
eigenstates of some operator $\hat{A}$. Measuring $\hat{A}$ without 
disturbing the wave function $\psi$ can be reduced to having a correlation
between the measured system with the coordinate ${\bf x}$ and the 
measuring apparatus with the coordinate ${\bf y}$, so that the total 
wave function takes the form 
\begin{equation}\label{wfy}                              
\Psi({\bf x},{\bf y},t)=\sum_a c_a\psi_a({\bf x},t)\phi_a({\bf y},t),
\end{equation}
where $\phi_a$ are also some orthonormal functions.
A realistic measurement lasts a finite time, so such a correlation 
must last for a finite time. 
Indeed, in the linear case, if $\phi_a$ are chosen such 
that the products $\psi_a\phi_a$ are solutions (when the interaction 
with the measuring apparatus is turned on), then 
the wave function (\ref{wfy}) is also a solution. On the other 
hand, if (\ref{wf}) is a solution when the interaction is turned off, 
it is typical of a nonlinear case that there will be no
solution of the form (\ref{wfy}) when the interaction is turned on.
Instead, the solution will have a form 
$\Psi=\sum_a \tilde{c}_a\psi_a\phi_a$, where $\tilde{c}_a\neq c_a$.  
This shows that {\em it is typically much more difficult to measure 
a quantity in nonlinear QM than that in ordinary linear QM}.

\subsection{Measurement for the classical Schr\"odinger equation}

After these general considerations of the problem of measurement 
in nonlinear QM, let us turn back to the case of nonlinear QM 
described by (\ref{schcl}). Do the results of the preceding 
subsection imply that, 
in practice, it is 
very difficult, if not impossible, to measure anything in the 
quantum world described by (\ref{schcl})? Fortunately, the answer is 
no! 

In the nonlinear case,
a linear combination of solutions does not need to be a solution, 
which, as we have seen, leads to problems for quantum measurements.
However, the particular form of the nonlinearity encoded in 
(\ref{schcl}) with (\ref{Q}) reveals that, in two general cases, 
a linear combination of solutions is also a solution.
First, if $\psi({\bf x},t)$ is a solution, then 
$c\psi({\bf x},t)$ is also a solution. Second, if some  
different solutions $\psi_a({\bf x},t)$ do not overlap, 
in the sense that 
\begin{equation}
\psi_a({\bf x},t)\psi_{a'}({\bf x},t)=0 \;\;\; {\rm for} 
\;\; a\neq a' ,
\end{equation}
then a linear combination of the form of (\ref{wf}) is also 
a solution. The first property shows that the overall normalisation 
of the solution is arbitrary, while the second property 
is a manifestation of locality.    
Note also that the eigenstates of the momentum operator
do overlap, while those of the position operator do not.
This indicates that the position can be easily measured, 
while the momentum cannot. 
However, to better understand what is measurable and what is not, 
below we formulate the theory in a slightly different form. 
    
Although $\psi$ does not satisfy a linear equation, there is a 
quantity that {\em does} satisfy a linear equation. This quantity is 
$R({\bf x},t)$ which satisfies the linear equation (\ref{lin}). Note 
that $R$ does not satisfy a linear equation in linear QM because 
there $R$ appears not only in (\ref{cons}) equivalent to (\ref{lin}), 
but also in (\ref{HJQ}). On the other hand, for the system described
by (\ref{schcl}), one can first solve Eq.~(\ref{HJ}) which does not contain 
$R$ and then insert the solution into (\ref{lin}) by considering $S$  
as a given external function. Thus we conclude that, {\em as far as the 
theory of measurement and effective collapse is concerned, the relevant
``wave function" associated with (\ref{schcl}) is not $\psi$, but $R$}.
As we show below,
this fact, together with the {\em positivity requirement} (\ref{R>0}), 
is the source of classical properties 
that emerge from the quantum theory described by (\ref{schcl}).
(Note, however, that it does not necessarily mean that the 
information contained
in $S$ is lost. Indeed, one can still consider the whole complex 
wave function $\psi$. In this case, in the effective collapse
of $\psi$, only $R$ changes, while $S$ remains the same.
More on this will be said in Sec.~\ref{PHSP}.)

Suppressing the time dependence of $R$ allows us to
introduce the bra and ket notation 
for states $R$, in complete analogy with ordinary QM.
Thus we have $R({\bf x})=\langle{\bf x}|R\rangle=\langle R|{\bf x}\rangle$,
where $|R\rangle$ and $\langle R|$ can be viewed as  
a column and row, respectively, representing the same real vector $R$
with the components $R({\bf x})$.   
We observe that the scalar product
\begin{equation}
\langle R_1| R_2\rangle \equiv \int d^3x 
\langle R_1|{\bf x}\rangle\langle{\bf x}| R_2\rangle
\equiv\int d^3x \, R_1({\bf x}) R_2({\bf x})
\end{equation}
is necessarily positive (i.e., real and nonnegative). 
We further note 
that the {\em only} complete orthogonal basis $\{ |R_i\rangle \}$
consistent with the positivity requirement 
$\langle {\bf x}|R_i\rangle\geq 0$ is the position basis 
$\{ |{\bf x}\rangle \}$. Thus, 
{\em the position basis is the preferred basis}. The most general 
state consistent with the positivity requirement is
\begin{equation}
|R\rangle = \int d^3x \, c({\bf x}) |{\bf x}\rangle ,
\end{equation}
where $c({\bf x})\geq 0$. 
In particular, there is no real state $R({\bf x})$ that is an 
eigenstate of the operator (\ref{p}), except for the trivial case 
in which the momentum eigenvalue is zero. Consequently, the state 
cannot ``collapse" to a nontrivial eigenstate of (\ref{p}), so
the operator $\hat{{\bf p}}$ cannot be 
(easily) measured. This implies that, in practice, {\em the 
Heisenberg uncertainty relation $\Delta x \Delta p \geq\hbar/2$
cannot be revealed by an experiment}. 

Note also that despite the fact that the operator $\hat{{\bf p}}$
cannot be measured, the momentum can still be measured 
in an {\em indirect} way.
In classical physics, one measures momentum 
by measuring two subsequent {\em positions} ${\bf x}_1$, ${\bf x}_2$
at times $t_1$, $t_2$, respectively, and defining the momentum as
${\bf p}=({\bf x}_2-{\bf x}_1)/m(t_2-t_1)$. This is how the  
momentum can be measured indirectly in the theory described by (\ref{schcl}).

\section{EMERGENCE OF CLASSICAL PHYSICS}
\label{CLAS}

\subsection{Classical statistics}

The origin 
of all nonclassical (i.e., typically quantum)
probabilistic phenomena of linear QM \cite{lal}
(such as those related to destructive interference,
EPR correlations, Bell inequalities, Hardy's paradox, etc.)
can be traced back to the fact that the scalar product
$\langle \psi_1| \psi_2\rangle$ between the probability amplitudes
$\psi_1$ and $\psi_2$ does not need to be positive.
Therefore, the positivity implies that there are no such
nonclassical probabilistic phenomena for (\ref{schcl}).

Similarly, for many-particle states, the particles 
can always be distinguished. For example, consider a symmetric
2-particle state   
$R({\bf x},{\bf y})=R_1({\bf x})R_2({\bf y})+R_2({\bf x})R_1({\bf y})$,
where $R_1$ and $R_2$ are {\em orthogonal}. In the
probability density $R^2$, the exchange term
$2R_1({\bf x})R_2({\bf x})R_1({\bf y})R_2({\bf y})$ vanishes, which,
just as in
ordinary QM, implies that the two particles can be regarded as
distinguishable.

Statistical physics can also be formulated in terms of density matrices.
A pure state
\begin{equation}\label{pures}
|R\rangle=\sum_{i} \sqrt{w_i} |R_i\rangle ,
\end{equation}
where $w_i \geq 0$, 
can be represented by a density matrix 
\begin{equation}
\hat{\rho}_{\rm pure}=|R\rangle\langle R| .
\end{equation}
It should be distinguished from the associated mixed state
\begin{equation}
\hat{\rho}_{\rm mix}=\sum_{i} w_i |R_i\rangle \langle R_i|  .
\end{equation}
One finds
\begin{equation}
\hat{\rho}_{\rm pure}=\hat{\rho}_{\rm mix} +
\sum_{i\neq j} \sqrt{w_i w_j} 
|R_i\rangle \langle R_j| .
\end{equation}
The mixed state is a diagonal state. The pure state is not diagonal, 
but its diagonal part is equal to the mixed state. 
Is there any measurable difference between the pure state and the 
mixed state? Since the operator $\hat{{\bf p}}$ is not measurable,
the most general measurable operator is of the form
$\hat{A}=A({\bf x})$. Since $|R_i\rangle$ are orthogonal, 
it follows that $\hat{A}$ is {\em diagonal}. Consequently, 
the average value of $\hat{A}$ in the mixed state is
\begin{equation}
\langle \hat{A}\rangle = {\rm Tr} (\hat{\rho}_{\rm mix} \hat{A})
= {\rm Tr} (\hat{\rho}_{\rm pure} \hat{A})
= \int d^3x\, \rho({\bf x}) A({\bf x}) ,
\end{equation}
where $\rho({\bf x})=R^2({\bf x})$ and 
$R({\bf x})=\sum_i \sqrt{w_i} R_i({\bf x})$.
This shows that {\em there is no 
measurable difference between pure states and the associated mixed states}.
These states differ only in the off-diagonal part, but
the positivity leads to the result that this part plays no measurable
role. Effectively, the off-diagonal part does 
not appear, which is a property of classical
statistical physics. 

The suppression of the off-diagonal part above is similar to, but still 
different from, the phenomenon of decoherence in ordinary QM \cite{zur,schl}.  
If an experiment in ordinary linear QM is designed such that 
the position basis plays a role of a preferred basis, 
then the decoherence induced by a practically unpredictable environment
leads to a classical-like statistics in which, for all 
practical purposes, one cannot distinguish between pure and mixed states.
However, the difference between QM and CM consists it the facts that 
(i) in CM one does not need a practically unpredictable environment to supress 
the off-diagonal terms, (ii) in QM the influence of the environment can be 
supressed, so that one can observe the effects of interference, 
which do not have a classical analog, 
and (iii) in QM the position basis does not 
necessarily need to be the preferred one, which also allows one to perform
measurements that can distinguish between CM and QM.  

One additional comment on the difference between classical and 
quantum statistics is in order.
Assume that some particular wave function $\psi$ satisfies both the 
quantum and the classical Schr\"odinger equation (in this case, $Q=0$).
Does it mean that the corresponding physical state satisfies both 
quantum and classical statistics? The answer is no! In fact, 
the statistics is not defined by a single state, but by the 
set of {\em all} possible states to which any given state may collapse.
This set is not defined by a particular solution, but by the 
{\em general} equation of motion -- quantum or classical Schr\"odinger
equation. (Recall that the superposition principle is crucial for the 
consistency of collapse, so in a classical collapse of $\psi$ only 
$R$ changes, while $S$ remains the same.)   
In this sense, in our approach one cannot say that the limit 
$Q\rightarrow 0$ corresponds to the classical or quantum limit 
of quantum or classical mechanics, respectively.
(See, however, Refs.~\cite{ghose,hall1,hall2} for different approaches.)

\subsection{Classical trajectories}

It remains to see why the measurements of positions at different times 
are consistent with the conventional classical picture according to which 
the particles move as given by (\ref{bohm}), despite the fact that 
we do not assume the validity of the ``hidden-variable" equation 
(\ref{bohm}). First, it is straightforward to see that the Ehrenfest theorem 
is valid for (\ref{schcl}): The average position is
\begin{equation}\label{Eh1}
\langle {\bf x}\rangle =\int d^3x \, \psi^*({\bf x},t) {\bf x} 
\psi({\bf x},t)=\int d^3x \, \rho({\bf x},t) {\bf x} .
\end{equation}
Using (\ref{Eh1}), (\ref{cons}), and (\ref{psi}) and integrating by 
parts, we find
\begin{equation}\label{Eh2}
\frac{d\langle {\bf x}\rangle}{dt} =\int d^3x \, \rho \frac{\nabla S}{m}
=\int d^3x \, \psi^* \frac{\hat{{\bf p}}}{m} \psi .
\end{equation} 
Similarly, using (\ref{Eh2}), (\ref{cons}), and (\ref{HJ}) and 
integrating by parts, we obtain
\begin{equation}\label{Eh3}
m\frac{d^2\langle {\bf x}\rangle}{dt^2} =\int d^3x \, \rho 
(-\nabla V)=\int d^3x \, \psi^* (-\nabla V)\psi .
\end{equation}
Thus, as in ordinary QM, the average position satisfies the classical 
equations of motion. However, the actual position may still be 
uncertain, so we still need to explain why particles appear
as pointlike in classical mechanics, despite the fact that 
we do not assume (\ref{bohm}). Of course, just as in ordinary QM, 
we can measure the position at a given time with arbitrary precision,
which corresponds to a ``collapse" of $R({\bf x})$ to an
arbitrarily narrow wave packet. However, in 
ordinary QM, even if the precision is perfect so that 
the wave packet is infinitesimally narrow at the given time, 
the wave packet will suffer dispersion, 
so that it will not be so narrow at later times. This is why particles 
in ordinary QM do not appear 
as pointlike objects with definite trajectories. 
On the other hand, the wave packets described by the nonlinear equation 
(\ref{schcl}) behave in an entirely different way. Eq.~(\ref{schcl})
contains infinitesimally narrow soliton solutions that do not suffer
dispersion. This can be seen as follows. Take an arbitrary 
{\em regular} solution $S$ of (\ref{HJ}) for which $\nabla S$ and
$\nabla^2 S$ are also regular and insert this solution into (\ref{cons}).
(Recall that one cannot do that in linear QM owing to the $R$-dependent 
$Q$-term in (\ref{HJQ}).) Consider a normalized Gaussian packet 
(not necessarily corresponding to a solution of (\ref{cons})) 
with the width $l$ 
\begin{equation}\label{Rl}
R_l({\bf x},t)=\pi^{-3/4}l^{-3/2} \exp [-({\bf x}-{\bf y}(t))^2/2l^2] ,
\end{equation}
where ${\bf y}(t)$ is an as yet undetermined function.
We see that $\lim_{l\rightarrow 0} \rho_l({\bf x},t)
\equiv \lim_{l\rightarrow 0} R^2_l({\bf x},t) = 
\delta^3({\bf x}-{\bf y}(t))$. By inserting (\ref{Rl}) into 
(\ref{cons}) and multiplying by $l$, we obtain
\begin{equation}\label{Rl2}
2\rho_l \frac{{\bf x}-{\bf y}}{l} \left( \frac{d{\bf y}}{dt} -
\frac{\nabla S}{m} \right) + l\rho_l \frac{\nabla^2 S}{m} =0.
\end{equation}
Now integrate (\ref{Rl2}) over $d^3x$ and consider the limit $l\rightarrow 0$.
Since $\nabla^2 S$ is regular, the last term in (\ref{Rl2}) 
does not contribute. In the first term we introduce the new integration
variable ${\bf z}=({\bf x}-{\bf y})/l$, so that equation (\ref{Rl2}) 
in the limit $l\rightarrow 0$ leads to
\begin{equation}
\frac{2}{\pi^{3/2}}
\int d^3z \, e^{-{\bf z}^2} {\bf z} \left( \frac{d{\bf y}}{dt} 
-\frac{\nabla S}{m} \right) =0.
\end{equation}
This shows that (\ref{Rl}) for $l\rightarrow 0$ {\em is} a solution of 
(\ref{lin}) if and only if
\begin{equation}\label{bohm'}
\frac{d{\bf y}(t)}{dt} =\frac{\nabla S({\bf y}(t),t)}{m} ,
\end{equation}     
which finishes our proof that 
(\ref{schcl}) contains infinitesimally narrow soliton 
solutions that do not suffer dispersion.
(Do not identify (\ref{bohm'}) with (\ref{bohm})! While 
$(\ref{bohm})$ describes hypothetical particle trajectories 
attributed to any state, (\ref{bohm'}) merely describes the 
motion of the crest of the wave packet (\ref{Rl}).)
Note that the derivation above may fail
for linear QM because, in this case, one cannot assume 
(without an explicit check) that
$\nabla^2 S$ is regular for $l\rightarrow 0$, so one cannot conclude
that the last term in (\ref{Rl2}) does not contribute for $l\rightarrow 0$.
Of course, for some special potentials $V$ in linear QM, it is still 
possible that $\nabla^2 S$ behaves such that the last term in 
(\ref{Rl2}) does not contribute for $l\rightarrow 0$. However, 
our argument above suggests that one should not expect this to be a 
generic feature of linear QM irrespective of $V$. Indeed, it is well 
known that most potentials $V$ in linear QM do not allow 
infinitesimally narrow non-dispersive solutions.

From the analysis above, we see that
{\em if a particle is measured to have a definite position 
at some time, then it will remain to have a definite position at later times
and this position will change with time according to the classical 
equations of motion}. 
However, in our interpretation, if the particle is {\em not} measured to have 
a definite position, then one is {\em not} allowed to claim 
that the particle {\em has} a definite position. While this interpretation 
contradicts the conventional interpretation of CM, it (just as in 
ordinary QM) does not contradict any measurable result of CM. 

\subsection{Classical phase space}
\label{PHSP}

The results of the preceding subsection show that there exist 
classical soliton states of the form
\begin{equation}\label{xp}
\psi_{\rm sol}({\bf x},t)=\sqrt{ \delta^3({\bf x}-{\bf y}(t)) } \,
e^{iS({\bf x},t)/\hbar} ,
\end{equation}
where ${\bf y}(t)$ is given by the integration of (\ref{bohm'}).
Although (\ref{xp}) is not an eigenstate of the operator 
$\hat{\bf p}$, at each time $t$ one can associate with this state 
a definite value of the momentum  
\begin{equation}\label{pphsp}
{\bf p}(t)=\nabla S({\bf y}(t),t) .
\end{equation}
The value (\ref{pphsp}) is obtained when the momentum is measured 
indirectly, by measuring two subsequent positions 
at $t$ and $t+dt$.
(In fact, since $\nabla \delta^3({\bf x}-{\bf y}(t))$ is weakly 
equal to zero, perhaps it also makes sense to say that 
$\hat{\bf p}\psi_{\rm sol}$ is weakly equal to ${\bf p}\psi_{\rm sol}$.)
Thus, at a fixed time, one can associate both the position 
${\bf y}$ and the momentum ${\bf p}$ with such a state, so, 
at a fixed time, we label such states as $\psi_{{\bf y},{\bf p}}$.
Such states correspond to points in the classical {\em phase} space.

Now consider a statistical ensemble of such states, 
$\{ w_{{\bf y},{\bf p}}, \psi_{{\bf y},{\bf p}} \}$, 
where $w_{{\bf y},{\bf p}}$ is the probability that the system 
is in the state $\psi_{{\bf y},{\bf p}}$. 
In general, such an ensemble cannot be described by a pure state 
$\psi$ or $R$. Instead, such an ensemble corresponds to the 
most general classical statistical ensemble in the phase space
and resembles the notion of mixed states in ordinary QM.
The average value of an observable $O({\bf x},{\bf p})$ is equal to
\begin{equation}
\langle O({\bf x},{\bf p}) \rangle = \int d^3x \int d^3p \,
w({\bf x},{\bf p}) O({\bf x},{\bf p}) ,
\end{equation}
where $w({\bf x},{\bf p}) \equiv w_{{\bf x},{\bf p}}$ and we 
assume that ${\bf x}$ and ${\bf p}$ are continuous labels.
As we demonstrate below, 
the general phase-space ensemble formulation contains the informations 
contained in the wave functions $\psi$ and $R$ as special cases.
  
As the first case, consider the case in which the initial information 
about the system is incoded in a pure state $\psi=R e^{iS/\hbar}$.
This corresponds to a phase-space statistical ensemble in which
\begin{equation}
w_{{\bf x},{\bf p}}(t)=R^2({\bf x},t) 
\delta^3({\bf p}-\nabla S({\bf x},t)) .
\end{equation}
If one measures ${\bf x}$ at $t$, then $\psi$ collapses to 
a wave function of the form of (\ref{xp}). This single measurement 
together with the initial information incoded in $\psi$ is 
sufficient to determine also the momentum ${\bf p}$.

As the second case, consider the case in which the initial information
about the system is incoded in a pure state $R$.
Now this corresponds to a phase-space statistical ensemble in which
$w_{{\bf x},{\bf p}}$ does not depend on ${\bf p}$, i.e.,
\begin{equation}
w_{{\bf x},{\bf p}}(t)=N R^2({\bf x},t), 
\end{equation} 
where $N=1/\int d^3p$ is the normalization factor.
If one measures ${\bf x}$ at $t$, then $R$ collapses to
a wave function of the form
$\sqrt{ \delta^3({\bf x}-{\bf y}(t)) }$, but the momentum ${\bf p}$
remains unknown. To determine ${\bf p}$, one must perform 
{\em two} measurements of ${\bf x}$, one at $t$ and the other at 
$t+dt$. Thus, two measurements are needed to determine 
both position and momentum incoded in the state $\psi_{{\bf x},{\bf p}}$.

In principle, both the first and the second case are physically possible, 
depending on the initial information about the system that 
an observer knows. However, in practical experimental 
situations, the second case is more often than the first one. 

\section{DISCUSSION}
\label{DISC}

In this paper, we have argued that
{\em all} measurable properties of classical mechanics 
can be predicted by the quantum theory based on the nonlinear 
Schr\"odinger equation (\ref{schcl}), without assuming the existence 
of particle trajectories satisfying (\ref{bohm}). 
Just as in ordinary linear QM, the only question 
that cannot be answered without this assumption is what causes
the effective wave-function collapse, i.e., what causes  
particles to take definite positions when $R({\bf x},t)$ is not a
$\delta$-function? If this question is irrelevant from the predictional 
point of view in linear QM, then, from the same point 
of view, it is also irrelevant in CM represented by the nonlinear 
Schr\"odinger equation (\ref{schcl}). Conversely, if 
one argues that this question is relevant in CM, so that
CM cannot be considered complete without Eq.~(\ref{bohm})
that explains the effective collapse, then
one can argue in the same way that 
linear QM also cannot be considered complete without (\ref{bohm}).
Of course, it is also consistent to adopt the conventional mixed 
interpretation in which CM is deterministic whereas QM is not, just as 
it is consistent to adopt a silly mixed interpretation in which QM is
deterministic whereas CM is not. However, it seems more reasonable to 
adopt a symmetric interpretation in which CM and QM are either both 
deterministic or both indeterministic. 
Nevertheless, we leave the choice 
of the interpretation to the reader.

At the end, we also note that our results raise interesting questions 
on more general nonlinear generalizations of QM. In general, {\em neither 
$\psi$ nor $R$ needs to satisfy a linear equation}. Does it really mean 
that nothing can be (easily) measured in such theories? Is it the 
reason why such theories do not describe the real world? Or is it 
the reason why we do not have any observational evidence for such 
theories? Or does it mean that our conclusion that linearity is crucial 
for practical measurability should be revised? 
Is it the Bohmian interpretation 
that needs to be adopted in order to have a sound interpretation
of more general nonlinear generalizations of QM? We leave the answers 
to these questions to the future research.   
 
\vspace{0.4cm}
\noindent
{\bf Acknowledgements.}
The author is grateful to anonymous referees for their 
objections that stimulated a more clear presentation.
This work was supported by the Ministry of Science and Technology of the
Republic of Croatia.

\end{document}